\documentclass[12pt]{article}

\usepackage{amsthm,graphicx,cite,braket}
\usepackage{epsf,latexsym}
\usepackage{amssymb,amsmath}
\usepackage{orcidlink}
\usepackage{times}
\usepackage{amscd}

\newcommand{\rmi}{{\rm i}}

\newcommand{\gr}{\Gamma _{\text{\tiny R}}}
\newcommand{\er}{E_{\text{\tiny R}}}

\begin{document}

\title{The Gamow Golden Rule of Multichannel Resonances}

\author{Rafael de la Madrid\,{\small $^{1,\dagger,\orcidlink{0000-0001-9350-5371}}$} \,
  Rodolfo~Id Betan\, {\small $^{2,3,*,\orcidlink{0000-0002-6813-3235}}$} \\
\small{$^{1}$Department of Physics, Lamar University, Beaumont, TX 77710,
  United States} \\
\small{$^{2}$ Instituto de F\'\i sica Rosario (CONICET-UNR), Ocampo y
    Esmeralda, Rosario 2000, Argentina} \\
  \small{ $^{3}$ Facultad de Ciencias Exactas, Ingenier\'\i a y Agrimensura
    (UNR), Av.~Pellegrini 250, Rosario 2000, Argentina} \\
    \small{$^\dagger$E-mail: \texttt{rafael.delamadrid@lamar.edu},
   $^*$E-mail: \texttt{idbetan@gmail.com}}}

\date{\small{February 26, 2026}}






            

\maketitle

\begin{abstract}
  \noindent
  We construct the Gamow Golden Rule of multichannel scattering, and use it
  to obtain the decay distributions, the partial decay constants, the
  partial decay widths, and the
  branching fractions of a resonance that has several decay modes. We
  exemplify the results using two coupled-channel square well potentials.
\end{abstract}


\section{Introduction}
\label{sec:intro}

Resonances are ubiquitous in quantum
mechanics~\cite{NEWTON,BAZ,TAYLOR,MAREK,RAKITYANSKY}. Experimentally,
they appear as sharp peaks in cross
sections~\cite{NEWTON,BAZ,TAYLOR,MAREK,RAKITYANSKY}, or as peaks in decay
(invariant mass) distributions~\cite{HIGGSA,HIGGSC,TETRA,CMS24,LHCb24}. The
standard theoretical tools to analyze resonant peaks in
cross sections are the analytical $S$-matrix and the R-matrix. The
standard tool to analyze decay distributions is the Golden Rule. The Golden
Rule describes decay distributions as a differential decay width, and
the total decay width as the integral of the differential decay width. In the
$S$-matrix formalism, the resonant energy is defined as a pole of the
$S$-matrix. In the Golden Rule, the resonant energy is obtained by solving
the Schr\"odinger equation under purely outgoing boundary conditions, which
lead to the same resonant energies as the poles of the $S$-matrix.

Although Fermi's Golden Rule has been very successful
in describing decay distributions, it is only an approximation
valid only for long-lived (sharp) resonances. In
Refs.~\cite{NPA15,NPA17,NPA18,NPA24},
we introduced the nonrelativistic Gamow Golden Rule, which is based on
the Gamow (resonant)
states, and which provides the exact expression for the Golden
Rule. The Fermi Golden Rule arises as an approximation to the
Gamow Golden Rule when the resonance is long-lived.

The Gamow Golden Rule introduced in Refs.~\cite{NPA15,NPA17,NPA18,NPA24} is
valid only for single-channel resonances. However, most experimental
resonances have several decay modes,
hence the need to generalize the Gamow Golden Rule to multichannel
resonances. In the present paper, we provide such generalization. We will
introduce the partial (channel) decay distributions, the partial decay
constants, the partial decay widths, and the branching fractions that
characterize the decay of a resonance into a particular channel. Because
many complicated multichannel systems are simplified by means of the
coupled-channel approximation, we will also
obtain the Gamow Golden Rule under such approximation. We will use
two coupled-channel square well
potentials~\cite{NEWTON,RAKITYANSKY,NEWTON61,GARRIDO,HAN,GROZDANOV} to
exemplify the quantities involved in the multichannel Golden Rule. All
our numerical results are obtained with Mathematica~\cite{MATHEMATICA}.

Although we exemplify them using the square barrier potential, 
our results apply to any spherically symmetric potential that is not
too singular at the origin and that decays fast enough at
infinity~\cite{TAYLOR,RAKITYANSKY}. In
particular, our results are valid for short range potentials. Charged
particles, whose Coulomb potential needs a special
treatment~\cite{TAYLOR,RAKITYANSKY}, are not described by our
formalism. However, just as in standard scattering
theory many results for neutral particles hold for charged
particles by replacing Bessel and Hankel functions with Coulomb
functions~\cite{TAYLOR,RAKITYANSKY}, it is likely that the Gamow Golden
Rule also holds for charge particles after the same replacement.

In the literature, there are two different types of boundary
conditions to obtain the bound states of a multichannel system, the
purely outgoing boundary condition (used for example in
Ref.~\cite{RAKITYANSKY}), and a mixture of scattering and outgoing
boundary conditions (used for example in Ref.~\cite{GARRIDO}). Different
boundary conditions lead to different bound states, and it is critical
to use the boundary conditions that are appropriate to describe
multichannel bound states. We will use the purely outgoing boundary
condition~\cite{RAKITYANSKY},
because it leads to the same bound and resonant energies as the poles of
the $S$-matrix.

Because they are eigenvectors of a linear operator, bound and resonant 
states are defined up to a multiplicative constant. Such constant is
determined by normalization of the natural square of the wave
function~\cite{BAZ,RAKITYANSKY,MAREK}. For single-channel bound states and
resonances, the expression for such normalization constant is well
known~\cite{BAZ,RAKITYANSKY,MAREK}. For multichannel bound states and
resonances, there are two different normalization constants in the
literature~\cite{BAZ,RAKITYANSKY}. Because different
normalizations yield different Golden Rules, it is
important to determine which one should be used. We will use a normalization
that is equivalent to that in Ref.~\cite{BAZ}.

The structure of the paper is as follows. In Sec.~\ref{sec:singlechannel}, we
review the most salient features of the Gamow Golden Rule in single-channel
scattering. In Sec.~\ref{sec:multichannel}, we construct the Gamow Golden Rule
of multichannel scattering by analogy with the single-channel case. We first
construct the Gamow Golden Rule in the momentum basis, and then in the
angular momentum basis for spherically symmetric potentials. In
Sec.~\ref{sec:coupledchannel}, we obtain the Gamow Golden Rule under the
coupled-channel approximation. In Sec.~\ref{sec:sphesymm-two}, we
solve the two-channel case for square well
potentials and obtain the explicit expressions of the bound states, resonant
states, and Green function. We will also show that multichannel
Gamow states can be normalized by factoring out the residue of the Green
function at the resonant pole as the product of two Gamow eigensolutions. In
Sec.~\ref{sec:calculation}, we present the numerical
calculation of the wave functions, decay energy spectra, partial decay
constants, partial decay widths, and branching fractions of the resonances
of the two-channel square well potential. The choice of the
potential parameters for our example is determined by two factors. First, those
parameters have been used before in the literature~\cite{HAN,GROZDANOV},
and second, they yield a resonance that is very close to a threshold. Many
resonances of great importance in astrophysics and
nuclear physics are situated close
to thresholds~\cite{RMP,PRC}, and our example shows
how the decay distribution provided by the Gamow Golden Rule is affected
by the presence of a threshold. In
Sec.~\ref{sec:conclusions}, we state our
conclusions. Appendices~\ref{app.a}, \ref{app.b} and~\ref{app.sphesymm-single}
contain supplementary calculations.

\section{The Single-Channel Gamow Golden Rule}
\setcounter{equation}{0}
\label{sec:singlechannel}

\subsection{The Gamow Golden Rule in the momentum basis}

 Let
$H=H_0+V$ be a Hamiltonian that produces resonances, where $H_0$ is the
free Hamiltonian and $V$ is the interaction potential. The eigenstates of 
the free Hamiltonian will be denoted by
$|\vec{p}\rangle$,
\begin{equation}
       H_0|\vec{p}\, \rangle = E_p |\vec{p} \, \rangle = \frac{p^2}{2m} 
                |\vec{p}\, \rangle \, .
         \label{freehe}
\end{equation}
The stationary states of scattering theory satisfy the Lippmann-Schwinger 
equation~\cite{TAYLOR,NEWTON,RAKITYANSKY},
\begin{equation}
       |\vec{p}\,^{+}\rangle = |\vec{p}\, \rangle + 
        \frac{1}{E-H_0+ i 0}V|\vec{p}\,^{+}\rangle \, .
         \label{lipsch}
\end{equation}
This equation is equivalent to the time-independent Schr\"odinger equation,
\begin{equation}
       H|\vec{p}\,^{+}\rangle = E_p |\vec{p}\, ^{+}\rangle = \frac{p^2}{2m} 
                |\vec{p}\, ^{+}\rangle \, ,
         \label{fulshdhe}
\end{equation}
subject to the boundary condition that, in the position representation, 
the ket $|\vec{p}\,^{+}\rangle$ represents an 
incoming plane wave and an outgoing spherical wave,
\begin{equation}
  \langle \vec{x}|\vec{p}\,^{+} \rangle \xrightarrow[r=|\vec{x}|\to \infty]{}
       (2\pi \hbar)^{-3/2}\left[ e^{i\vec{p}\cdot\vec{x}/\hbar}+
            f(\vec{p} \to p\hat{x} )
                        \frac{e^{ipr/\hbar}}{r}\right]  \, , 
          \label{asymptotic+}
\end{equation}
where $f(\vec{p} \to p\hat{x} )$ is the scattering amplitude.

We will denote the resonant (Gamow) state by $|p_{\rm res} \rangle$, where 
$\frac{p_{\rm res}^2}{2m} = E_{\rm res} = \er - i \gr /2$ is the complex 
resonant energy of a 
decaying state. Instead of the Lippmann-Schwinger equation~(\ref{lipsch}),
the resonant states satisfy~\cite{WOLF,HERMON}
\begin{equation}
    |p_{\rm res}\rangle = 
           \frac{1}{E_{\rm res} -H_0 +\rmi 0}V |p_{\rm res}\rangle \, .
         \label{inteGam0}
\end{equation} 
This integral equation is equivalent to the time-independent 
Schr\"odinger equation
\begin{equation}
       H|p_{\rm res}\rangle = E_{\rm res} |p_{\rm res} \rangle
       = \frac{p_{\rm res}^2}{2m} |p_{\rm res} \rangle \, ,
       \label{timeindschgs}
\end{equation}
subject to a purely outgoing boundary condition,
\begin{equation}
  \langle \vec{x}|p_{\rm res} \rangle \xrightarrow[|\vec{x}|\to \infty]{}
                      g(\hat{x},p_{\rm res})  \frac{e^{ip_{\rm res}r/\hbar}}{r} \, ,
       \label{pobc}
\end{equation}
where the function $g(\hat{x},p_{\rm res})$ does not depend on $r$.

The single-channel Gamow Golden Rule can
be derived very easily~\cite{NPA24}. First, we 
take the inner product of Eq.~(\ref{inteGam0}) with $\langle \vec{p}|$ and
use Eq.~(\ref{freehe}) to obtain
\begin{equation}
   \langle \vec{p} |p_{\rm res}\rangle = 
           \langle \vec{p}|\frac{1}{E_{\rm res} -H_0 +\rmi 0}V |p_{\rm res}\rangle 
= \frac{1}{E_{\rm res} -E_p} \langle \vec{p}|V |p_{\rm res}\rangle \, . 
         \label{inteGamp0}
\end{equation}
The decay spectrum $d\Gamma/d^3p$ of a
resonance decaying into a free particle of 
momentum $\vec{p}$ is simply the absolute value squared of
Eq.~(\ref{inteGamp0}),
\begin{equation}
   \frac{d\Gamma (p_{\rm res} \to \vec{p})}{d^3p} = 
  |\langle \vec{p} |p_{\rm res}\rangle|^2 =
\frac{1}{(E_p-\er)^2+(\gr/2)^2}  
         |\langle \vec{p} |V|p_{\rm res}\rangle|^2 \, .
    \label{momenproden2}
\end{equation}
The total decay constant for decay into the continuum is obtained by integrating
Eq.~(\ref{momenproden2}),
\begin{equation}
    \Gamma = \int d^3p \, \frac{1}{(E_p-\er)^2+(\gr/2)^2}  
    |\langle \vec{p} |V|p_{\rm res}\rangle|^2 \, . 
    \label{momenproden2bto}
\end{equation}

As shown in Ref.~\cite{NPA24}, one can also define differential and 
total decay widths as
\begin{equation}
       \frac{d\overline{\Gamma}(E)}{dE} = \gr \frac{d\Gamma(E)}{dE} \, ,
         \qquad 
        \overline{\Gamma} = \gr \Gamma \, .
           \label{dewvsdctb} 
\end{equation}
Because in general $\Gamma \neq 1$, $\overline{\Gamma}$ is different
from $\gr$.

\subsection{The Gamow Golden Rule in the angular momentum basis}

When the interaction is spherically symmetric, we can obtain the Gamow Golden
Rule in the angular momentum basis by using an argument that
is almost identical to the one in the previous section~\cite{NPA24}. The
resulting $l$th partial-wave differential decay
constant for decay of a resonance of angular momentum quantum numbers $l,m$ into
a free particle of angular momentum quantum numbers $l',m'$ is given by
\begin{equation}
\frac{d\Gamma (E_{\rm res}, l,m \to E,l',m')}{dE} =  
           \frac{1}{(E-\er)^2+(\gr/2)^2} |\langle E|V |E_{\rm res} \rangle _l|^2 
            \delta_{l,l'}\delta_{m,m'}   \, ,            
         \label{inteGam3}
\end{equation}
where the reduced matrix element of the interaction is given by
\begin{equation}
        \langle E|V|E_{\rm res}\rangle _l
         =
          \int_0^{\infty} dr\, \psi_{l}^0(r,E_p)^* V(r) u_l(r;E_{\rm res}) \, ,
         \label{inteGam2}
\end{equation}
with $u_l(r;E_{\rm res})$ being the radial part of the Gamow state,
\begin{equation}
         \langle \vec{x}|E_{\rm res},l,m \rangle =
           \frac{u_l(r;E_{\rm res})}{r}Y_l^m(\hat{x} ) \, ,
\end{equation}
and $\psi_{l}^0(r,E_p)$ being the radial part of the free wave
function~\cite{TAYLOR},
\begin{equation}
       \langle \vec{x}|E,l,m\rangle = 
                \frac{\psi_l^0(r;E_p)}{r} Y_l^m(\hat{x}) =
      i^l  
         \sqrt{\frac{2m}{\pi \hbar p}} \,
                \hat{j}_l(pr/\hbar) \frac{1}{r} Y_l^m(\hat{x})  \, ,
        \label{angulmombdefree}
\end{equation}
where $\hat{j}_l(x)$ are the Riccati-Hankel
functions~\cite{TAYLOR}. The $l$th partial-wave decay constant is obtained
by integration of Eq.~(\ref{inteGam3}) over the scattering spectrum,
\begin{equation}
       \Gamma_l  = \int dE
       \frac{1}{(E-\er)^2+(\gr/2)^2} |\langle E|V |E_{\rm res} \rangle _l|^2 \, .
           \label{inteGam4}
\end{equation}

\section{The Multichannel Gamow Golden Rule}
\setcounter{equation}{0}
\label{sec:multichannel}

\subsection{The multichannel Gamow Golden Rule in the momentum basis}

We are going to use the notation of Ref.~\cite{TAYLOR} and write
the Lippmann-Schwinger equation
for channel $\alpha$ in terms of the appropriate
relative momenta $\underline{p}$ as 
\begin{equation}
       |\underline{p},\alpha^{+}\rangle =
       |\underline{p},\alpha \rangle + \frac{1}{E-{H^{\alpha}}+ i0}
          V^{\alpha} |\underline{p},\alpha^{+}\rangle \, , 
       \label{lsmulticmom}
\end{equation}
where $H^{\alpha}$ ($V^{\alpha}$) is the ``free'' $\alpha$-channel 
Hamiltonian (potential), $H=H^{\alpha}+V^{\alpha}$
is the total Hamiltonian, and $|\underline{p},\alpha \rangle$ are eigenkets
of the ``free'' $\alpha$-channel Hamiltonian,
\begin{equation}
       H^{\alpha}|\underline{p},\alpha \rangle = 
       E |\underline{p},\alpha \rangle
       = (E_{{\underline{p_\alpha}}} +E_{{\rm th},\alpha})
         |\underline{p},\alpha \rangle
       \, ,
      \label{mcfreehald}
\end{equation}
where we have split the energy $E$ into a kinetic part
$E_{{\underline{p_\alpha}}}$ and a threshold part
$E_{{\rm th},\alpha}$. Equation~(\ref{lsmulticmom}) implies that 
$|\underline{p},\alpha^{+}\rangle$ 
are eigenkets of the total Hamiltonian subject to the boundary condition 
of an incident plane wave and outgoing 
spherical waves. As in the single-channel
case, the second term on the right-hand side of Eq.~(\ref{lsmulticmom})
produces the outgoing spherical wave.

By analogy with the single-channel case, we are going to assume that
the Gamow state describing a multichannel resonance that decays into
channel $\alpha$ is an eigenfunction of the Hamiltonian
subject to a purely outgoing boundary condition in channel $\alpha$, and
therefore satisfies the following integral equation:
\begin{equation}
       |{\underline{p}}_{\rm res}\rangle = \frac{1}{E_{\rm res}-H^{\alpha}+ \rmi 0}
          V^{\alpha}| \underline{p}_{\rm res} \rangle \, . 
       \label{lsmulticgmom}
\end{equation}
Equation~(\ref{lsmulticgmom}) should be satisfied for any channel 
into which the resonance can decay. The analogy with the single-particle 
case is clear. Formally, the Lippmann-Schwinger equation for a 
single-channel resonance, Eq.~(\ref{inteGam0}), is obtained from the
single-channel Lippmann-Schwinger equation, Eq.~(\ref{lipsch}),
by omitting the free incoming plane wave $|\vec{p}\rangle$. Similarly,
the Lippmann-Schwinger equation for a 
multichannel resonance, Eq.~(\ref{lsmulticgmom}), is obtained from the
multichannel Lippmann-Schwinger equation, Eq.~(\ref{lsmulticmom}),
by omitting the free incoming plane wave $|\underline{p},\alpha \rangle$. 

If we take the inner product of 
Eq.~(\ref{lsmulticgmom}) with $\langle \underline{p},\alpha|$ and
use Eq.~(\ref{mcfreehald}), we obtain
\begin{equation}
      \langle \underline{p}, \alpha |\underline{p}_{\rm res}\rangle = 
         \langle \underline{p}, 
           {\rm \alpha}|\frac{1}{E_{\rm res}-H^{\alpha}+\rmi 0}
          V^{\alpha}| \underline{p}_{\rm res} \rangle =
        \frac{1}{E_{\rm res}-E} 
        \langle \underline{p}, {\rm \alpha}|V^{\alpha}| 
                 \underline{p}_{\rm res} \rangle  \, . 
       \label{lsmulticgmom1}
\end{equation}
The absolute value squared of this quantity is the probability density
(which we identify with the partial differential decay constant) that a 
resonance decays into channel $\alpha$ with momentum $\underline{p}$,
\begin{equation}
      \frac{d\Gamma_{\alpha} (\underline{p}_{\rm res}\to 
                \underline{p},\alpha)}{d\underline{p}} =
       \frac{1}{(E -\er)^2+ (\gr /2)^2} 
        |\langle \underline{p},{\rm \alpha}|V^{\alpha}| 
              \underline{p}_{\rm res} \rangle|^2  \, . 
       \label{lsmulticgmom2}
\end{equation}
Similarly to the single-channel case, we can interpret this equation by
saying that the probability per unit of momentum volume for a resonance
to decay into channel $\alpha$ is given by the Lorentzian lineshape
times the matrix element of the $\alpha$-channel interaction.

The partial decay constant for channel $\alpha$ is obtained by integrating
Eq.~(\ref{lsmulticgmom2}),
\begin{equation}
       \Gamma _{\alpha} = \int d\underline{p} \, 
     \frac{d\Gamma_{\alpha} (\underline{p}_{\rm res}\to 
                \underline{p};\alpha)}{d\underline{p}} =
        \int d\underline{p} \,
         \frac{1}{(E -\er)^2+ (\gr /2)^2} 
        |\langle \underline{p};{\rm \alpha}|V^{\alpha}| 
              \underline{p}_{\rm res} \rangle|^2  \, ,
           \label{minteGam4}
\end{equation} 
and the total decay constant is 
\begin{equation}
       \Gamma =\sum_{\alpha} \Gamma _{\alpha}  \, .
\end{equation}
As in the single-channel case, we can define the partial and the total
decay widths as
\begin{equation}
  \overline{\Gamma}_{\alpha}= \Gamma_{\alpha} \gr \, , \qquad
  \overline{\Gamma}  = \sum_{\alpha} \overline{\Gamma}_{\alpha}=\Gamma \gr \, .
\end{equation}
The branching fractions for decay into each channel are defined as
\begin{equation}
       {\cal B}_{\alpha}=
       {\cal B} (E_{\rm res} \to \alpha ) = \frac{\Gamma _{\alpha}}{\Gamma}
       = \frac{\overline{\Gamma}_{\alpha}}{\overline{\Gamma}}\, .
\end{equation}

\subsection{The multichannel Gamow Golden Rule in the angular-momentum basis}

For spherically symmetric interactions, and if we consider only
two-body channels of spinless particles, we can work with an angular
momentum basis and obtain equations that are analogous to the single-channel
case. Using 
an angular momentum basis, the Lippmann-Schwinger equation of a 
multichannel Gamow state is
\begin{equation}
       |E_{\rm res},l,m \rangle = \frac{1}{E_{\rm res}-H^{\alpha}+ \rmi 0}
          V^{\alpha}| E_{\rm res}, l, m \rangle \, . 
       \label{lsmulticgen}
\end{equation}
By taking the inner product of this equation with the eigenbra
$\langle E_{\underline{p}}^{\alpha},l_{\alpha},m_{\alpha}|$ of the ``free''
$\alpha$-channel
Hamiltonian, where $l_{\alpha}$ and $m_{\alpha}$ denote the collection of 
angular-momentum quantum numbers of the stable decay products,
and by using the fact that $\langle E_{\underline{p}}^{\alpha},l_{\alpha},m_{\alpha}|$ 
is an eigenvector of the ``free'' $\alpha$-channel Hamiltonian $H^{\alpha}$
with eigenvalue $E$, we get
\begin{eqnarray}
       \langle E_{\underline{p}}^{\alpha},l_{\alpha},m_{\alpha} |E_{\rm res},l,m \rangle &=& 
       \langle E_{\underline{p}}^{\alpha},l_{\alpha},m_{\alpha}|\frac{1}{E_{\rm res}-H^{\alpha}+ \rmi 0}
          V^{\alpha} |E_{\rm res},l,m \rangle \nonumber  \\
  & =& \frac{1}{E_{\rm res}-E} \langle E_{\underline{p}}^{\alpha},l_{\alpha},m_{\alpha} |V^{\alpha} |E_{\rm res},l,m\rangle \, .
       \label{lsmulticg2}
\end{eqnarray}
Since $\langle E_{\underline{p}}^{\alpha},l_{\alpha},m_{\alpha} |E_{\rm res},l,m\rangle$
is the probability amplitude for the decay of the resonance into a stable
state of channel $\alpha$, the differential $l$th partial-wave decay constant
for channel $\alpha$ can be written as follows:
\begin{equation}
      \frac{d\Gamma _{\alpha}(E_{\rm res},l,m \to E_{\underline{p}}^{\alpha},l_{\alpha},m_{\alpha})}{dE_{\underline{p}}^{\alpha}} =  
           \frac{1}{(E-\er)^2+(\gr/2)^2} 
       |\langle E_{\underline{p}}^{\alpha},l_{\alpha},m_{\alpha}|V^{\! \alpha} |E_{\rm res},l,m \rangle |^2 
           \, .     
\label{lsmulticg4}
\end{equation}
We can interpret this equation by saying that a resonance (represented by
$|E_{\rm res},l,m\rangle$) decays into a stable state of channel $\alpha$ 
(represented by
$|E_{\underline{p}}^{\alpha},l_{\alpha},m_{\alpha} \rangle$) through the coupling provided by the channel
potential $V^{\alpha}$. Similarly to the single-channel case, the decay 
energy spectrum of a resonance into 
the continuum of channel $\alpha$ is given by the Lorentzian lineshape times 
the matrix element of the interaction of channel $\alpha$.

\section{The Gamow Golden Rule in the Coupled-Channel Approximation}
\setcounter{equation}{0}
\label{sec:coupledchannel}

\subsection{The coupled-channel approximation of the Gamow Golden Rule in 
in the momentum basis}

For concreteness, we will follow Ref.~\cite{TAYLOR} and
consider the specific example of particle $a$ impinging on the ground state 
$(bc)_1\equiv \eta _1(\vec{x}_{\rm b})$ of a target compound made of two
particles $b$ and $c$, where $c$ is a heavy, fixed particle. The coordinates
of the projectile and target will be denoted by $\vec{x}$ and
$\vec{x}_{\rm tar}$, respectively.

In the coupled-channel approximation, one starts with the expansion
of the stationary states in terms of target states~\cite{TAYLOR},
\begin{equation}
        \psi _{\rm res}(\vec{x},\vec{x}_{\rm tar}) =
          \sum_{\alpha =1}^N \phi^{\rm res}_{\alpha}(\vec{x})\, 
            \eta_{\alpha}(\vec{x}_{\rm tar})
          \, ,
    \label{targetexpnsres}
\end{equation}
where we have assumed that there are $N$ target states $\eta_{\alpha}$,
and where
\begin{equation}
         H = H^1+V^1 = \frac{\hat{P}^2}{2m_a}+\hat{H}_{\rm tar} +V^1 \, ,
\end{equation}
and
\begin{equation}
       \hat{H}_{\rm tar}\eta _{\alpha} =E_{{\rm th},\alpha} \eta _{\alpha}  \, .
\end{equation}

In the position representation, the eigenket $|\vec{p}_\alpha \rangle$ of 
$H^{\alpha}$ is given by
\begin{equation}
        \langle \vec{x},\vec{x}_{\rm tar} | \vec{p}_\alpha \rangle =
            \frac{e^{i\vec{p}_{\alpha}\cdot \vec{x}/\hbar}}{(2\pi \hbar)^{3/2}}
             \eta_{\alpha}(\vec{x}_{\rm tar}) =
             \phi_{\alpha}^{\rm free}(\vec{x}) \eta_{\alpha}(\vec{x}_{\rm tar}) \, ,
             \label{alphafreetar}
\end{equation}
which describes a free particle of reduced mass $\mu_{\alpha}$ and momentum
$p_{\alpha}= \sqrt{2\mu_{\alpha} (E-E_{{\rm th},\alpha})}$,
and a target state of wave function $\eta _{\alpha}(\vec{x}_{\rm tar})$. By
inserting the expansion~(\ref{targetexpnsres}) into the
matrix element of the interaction, we obtain
\begin{equation}
  \langle \vec{p}_\alpha|V^1|p_{\rm res}\rangle
               =  \sum_{\beta =1}^N
   \langle\phi _{\alpha} ^{\rm free}(\vec{x})|\bar{V}_{\alpha,\beta}(\vec{x})| 
\phi ^{\rm res}_{\beta}(\vec{x}) 
      \rangle \, , 
\end{equation}
where 
$\bar{V}_{\alpha,\beta}(\vec{x})= \langle \eta_{\alpha}(\vec{x}_{\rm target})| 
             V^1(\vec{x},\vec{x}_{\rm target})|\eta_{\beta}(\vec{x}_{\rm target})\rangle$ are the potentials that result from integrating
the target coordinates out. Hence, in the coupled-channel approximation, 
Eq.~(\ref{lsmulticgmom2}) becomes
\begin{equation}
      \frac{d\Gamma_{\alpha} (p_{\rm res}\to \vec{p}_\alpha)}{d^3p_\alpha } =
       \frac{1}{(E - \er)^2+ (\gr /2)^2} 
          \left| \sum _{\beta =1}^N
   \langle\phi _{\alpha} ^{\rm free}(\vec{x})|\bar{V}_{\alpha,\beta}(\vec{x})| 
         \phi ^{\rm res}_{\beta}(\vec{x})\rangle \right| ^2
          . 
       \label{lsmulticg5}
\end{equation}
It is important to keep in mind that in Eq.~(\ref{lsmulticg5}), $E$ runs
from the threshold energy $E_{{\rm th},\alpha}$ to infinity, whereas
\begin{equation}
            E_{p_{\alpha}}=\frac{p_{\alpha}^2}{2\mu_{\alpha}} = E-E_{{\rm th},\alpha}
   \label{epaeandeth}
\end{equation}
runs from $0$ to infinity.

\subsection{The coupled-channel approximation of the Gamow Golden Rule in 
in the angular momentum basis}

When the potentials are spherically symmetric,
$\bar{V}_{\alpha,\beta} (\vec{x})=\bar{V}_{\alpha,\beta} (r)$, we can express the
multichannel Gamow Golden Rule in an angular momentum basis:
\begin{equation}
      \frac{d\Gamma_{\alpha,l}}{dE_{p_\alpha}} =
       \frac{1}{(E - \er)^2+ (\gr /2)^2} 
          \left| \sum _{\beta =1}^N
       \langle E_{p_{\alpha}}|\bar{V}_{\alpha,\beta}|E_{\rm res}, \beta \rangle _l
                 \right| ^2   \, ,      
       \label{lsmulticg5ss}
\end{equation}
where $\langle r| E_{p_{\alpha}}\rangle$ ($\langle r |E_{\rm res}, \beta \rangle$)
is the radial component of the free (resonant) wave function, see
Eqs.~(\ref{gamowparsochannel}) and~(\ref{freeparsoechannel}) below. The
proof is similar to the single-channel case~\cite{NPA24}, and can be found
in Appendix~\ref{app.a}. The $l$-th partial-wave decay constants are given
by
\begin{equation}
      \Gamma_{\alpha,l}= \int_0^{\infty} dE_{p_\alpha} \, 
       \frac{1}{(E - \er)^2+ (\gr /2)^2} 
          \left| \sum _{\beta =1}^N
       \langle E_{p_{\alpha}}|\bar{V}_{\alpha,\beta}|E_{\rm res}, \beta \rangle _l
                 \right| ^2        
       \label{lsmulticg5ss2}
\end{equation}
The $l$-th partial-wave decay widths and branching fractions are
\begin{equation}
  \overline{\Gamma}_{\alpha,l} = \Gamma _{\alpha,l} \gr \, , 
              \qquad
         \overline{\Gamma}_{l} = \sum_{\alpha} \overline{\Gamma}_{\alpha,l} 
           \, , \qquad 
  {\cal B}_{\alpha,l} = \frac{\Gamma _{\alpha,l}}{\Gamma _l} =
            \frac{\overline{\Gamma}_{\alpha,l}}{\overline{\Gamma _l}} \, .
           \label{bflth}
\end{equation}

The derivation of Eq.~(\ref{lsmulticg5ss}) in Appendix~\ref{app.a} ultimately
relies upon general
considerations on the asymptotic behavior of
the multichannel Lippmann-Schwinger equation. To double check our results,
in Appendix~\ref{app.b} we provide an alternative derivation of
Eq.~(\ref{lsmulticg5ss}) using the explicit
expression of the coupled-channel Schr\"odinger equation.

It is enlightening to write down explicitly the decay energy
spectrum~(\ref{lsmulticg5ss}) for a two-channel system:
\begin{eqnarray}
      \frac{d\Gamma _{1,l}}{dE_{p_1}} =  
           \frac{1}{(E_{p_1}+E_{{\rm th},1} -\er)^2+(\gr/2)^2} 
       \left|
   \langle E_{p_1}|\bar{V}_{11}|E_{\rm res},1\rangle_l +
       \langle E_{p_1}|\bar{V}_{12}| E_{\rm res},2 \rangle_l \right| ^2
            , 
           \label{lsmulticg91}    \\
\frac{d\Gamma _{2,l}}{dE_{p_2}} =  
           \frac{1}{(E_{p_2}+E_{{\rm th},2} -\er)^2+(\gr/2)^2} 
       \left|
   \langle E_{p_2}|\bar{V}_{21}| E_{\rm res},1\rangle_l +
       \langle E_{p_2}|\bar{V}_{22}|E_{\rm res},2 \rangle _l \right| ^2
            ,
        \label{lsmulticg92}
\end{eqnarray}
where we have used Eq.~(\ref{epaeandeth}) to write
$E=E_{p_{\alpha}}+E_{{\rm th},\alpha}$, where
$|E_{\rm res},1 \rangle$ and $|E_{\rm res},2 \rangle$ are the channel
components of the Gamow state,
\begin{equation}
\braket{r|E_{\rm res}} =  \left( \begin{array}{c}
            \braket{r|E_{\rm res}, 1} \\ [1ex]
            \braket{r|E_{\rm res},2}                     
  \end{array}
  \right) = \left( \begin{array}{c}
            u_1(r;E_{\rm res}) \\ [1ex]
            u_2(r;E_{\rm res})                     
  \end{array}
  \right)  ,
     \label{gamowparsochannel}
\end{equation}
and where
\begin{equation}
  \braket{r|E_{p_1}} =
     \left( \begin{array}{c}
            \psi ^0_l(r;E_{p_1}) \\ [1ex]
                                     0
  \end{array}
  \right) , \qquad
  \braket{r|E_{p_2}}  =
  \left( \begin{array}{c}
    0 \\ [1ex]
      \psi ^0_l(r;E_{p_2})
  \end{array}
  \right) 
    \label{freeparsoechannel}
\end{equation}
are the column vectors that represent a free particle with
angular momentum $l$ and linear momentum $p_1$ and $p_2$ in channels~1 and~2,
respectively, with $\psi ^0_l(r;p)$ given in Eq.~(\ref{angulmombdefree}). From
Eqs.~(\ref{lsmulticg91}) and~(\ref{lsmulticg92}) it is
clear that the decay distribution of a resonance into a particular channel
depends on the interference of the components of the Gamow state along all
the channels to which the resonance can decay, a manifestation of the
coupling between the channels.

\section{The Two-Channel Square Well Potential}
\setcounter{equation}{0}
\label{sec:sphesymm-two}

In this section, we exemplify the above formalism using two coupled-channel
square well potentials. Since in the literature there are
two different solutions for the bound states of this
system~\cite{RAKITYANSKY,GARRIDO}, and since there is no explicit, fully
analytic solution for the resonant states, we are going to present a detailed,
systematic solution for this potential. The upshot is that the method of
solution of the two-channel case is completely analogous to that of the
single-channel case, which we include in Appendix~\ref{app.sphesymm-single}
for the sake of completeness.

\subsection{The Schr\"odinger equation and its general solution}

When the potentials are spherically symmetric,
the $s$-wave radial parts of the stationary states satisfy the following
coupled radial equations~\cite{TAYLOR}:
\begin{eqnarray}
  -\frac{\hbar ^2}{2\mu_1} \frac{d^2\psi_1}{dr^2} + \bar{V}_{11}(r) \psi_1(r)
  +\bar{V}_{12}(r) \psi_2(r) = (E-E_{\rm th,1}) \psi_1(r) \, ,
     \label{channel1a} \\
      -\frac{\hbar ^2}{2\mu_2} \frac{d^2\psi_2}{dr^2} + \bar{V}_{21}(r) \psi_1(r)
      +\bar{V}_{22}(r) \psi_2(r) = (E-E_{\rm th,2}) \psi_2(r) \, .
      \label{channel2a}
\end{eqnarray}
If we define the channel wave numbers and potentials as
\begin{equation}
  k_{\alpha} = \sqrt{ \frac{2\mu_{\alpha}}{\hbar ^2} (E-E_{{\rm th,}\alpha})}  \, ,
 \qquad
   V_{\alpha \beta}(r) = \frac{2\mu_\alpha}{\hbar^2}  \bar{V}_{\alpha \beta}(r) \, ,
\end{equation}
then we can write Eqs.~(\ref{channel1a}) and~(\ref{channel2a})
as
\begin{eqnarray}
  - \frac{d^2\psi_1}{dr^2} + V_{11}(r)
     \psi_1(r)
     + V_{12}(r) \psi_2(r) =
      k_1^2 \psi_1(r) \label{channel1c} \\
      - \frac{d^2\psi_2}{dr^2} + V_{21}(r) \psi_1(r)
         + V_{22}(r) \psi_2(r) = k_2^2 \psi_2(r)
           \label{channel2c}
\end{eqnarray} 
Let us define the column wave function, wave number matrix and potential
matrix as
\begin{equation}
  {\sf \Psi} = \left( \begin{array}{c}
    \psi_1 \\
    \psi_2
  \end{array}
  \right) \, , \quad 
   {\sf K} =  \left( \begin{array}{cc}
    k_1 & 0 \\
    0    & k_2
  \end{array}
   \right)  \, , \quad 
    {\sf V} =  \left( \begin{array}{cc}
    V_{11} & V_{12} \\
    V_{21} & V_{22}
  \end{array}
      \right)  \, . 
         \label{potentmatrix}
\end{equation}
We may sometimes specify the position and energy/wave number dependence of
${\sf \Psi}$ as
${\sf \Psi}\equiv {\sf \Psi}(r;E)\equiv {\sf \Psi}(r;\vec{k})\equiv
{\sf \Psi}(r; k_1,k_2)$. The system of Eqs.~(\ref{channel1c})
and~(\ref{channel2c}) can be written as
\begin{equation}
  -{\sf \Psi}'' + {\sf V} {\sf \Psi} ={\sf K}^2 {\sf \Psi} \, .
    \label{cseqmformagen} 
\end{equation}
The entries of the potential matrix~(\ref{potentmatrix}) are square
barrier/well potentials of heights/depths $V_{\alpha \beta}$ and the
same range $a$:
\begin{equation}
  V_{\alpha \beta}(r) =
  \left\{ \begin{array}{cc}
    V_{\alpha \beta}  &  r<a \, ,  \\
    0      & r>a \, .
  \end{array}
  \right.
    \label{muchswelpot}
\end{equation}
For the sake of simplicity, we are going to assume that
the reduced mass is the same for all channels, $\mu_1=\mu_2=\mu$. This
will ensure that ${\sf V}$ is a symmetric matrix.

When the potentials are given by Eq.~(\ref{muchswelpot}),
Eq.~(\ref{cseqmformagen}) can be written as
\begin{eqnarray}
  {\sf \Psi}'' + \left[{\sf K}^2-{\sf V}\right] {\sf \Psi} =0
              \, , &  & r<a \, ,
  \label{cseqmforma} \\
   {\sf \Psi}'' + {\sf K}^2 {\sf \Psi} =0 \, , & & r>a \, .
     \label{cseqmformb}
\end{eqnarray}
Equation~(\ref{cseqmformb}) can be easily solved because the equations
are uncoupled when $r>a$. To solve Eq.~(\ref{cseqmforma}),
we are going to obtain the eigenvalues of the matrix
${\sf K}^2-{\sf V}$~\cite{NEWTON61,NEWTON,GARRIDO}. A
straightforward calculation shows that the eigenvalues are
\begin{equation}
  q_{\pm}^2= \frac{1}{2} \left[ (k_1^2-V_{11})+ (k_2^2-V_{22}) \pm
   \sqrt{ \left[ (k_1^2-V_{11}) - (k_2^2-V_{22})\right]^2 +4 V_{12}V_{21}}
    \, \right] \, .
\end{equation}
The corresponding normalized eigenvectors are
\begin{equation}
  |q_+^2\rangle =\frac{1}{N} \left( \begin{array}{c}
    V_{12} \\ [1ex]
    k_1^2-V_{11}-q_+^2
  \end{array}  \right) \, ,
  \qquad
  |q_-^2\rangle = \frac{1}{N} \left( \begin{array}{c}
    k_2^2-V_{22}-q_-^2 \\ [1ex]
    V_{21}
  \end{array}  \right) \, ,
     \label{eigenvectors}
\end{equation}
where the normalization constant $N$ is given by
\begin{equation}
  N^2= V_{12}^2 + (k_1^2-V_{11}-q_+^2)^2 = V_{21}^2 + (k_2^2-V_{22}-q_-^2)^2 \, .
\end{equation}
We can use the eigenvectors~(\ref{eigenvectors}) to obtain the change of
basis matrix ${\sf O}$ that diagonalizes the matrix ${\sf K}^2-{\sf V}$:
\begin{equation}
  {\sf O} = \frac{1}{N} \left(  \begin{array}{ccc}
                 V_{12} & & k_1^2-V_{11}-q_+^2 \\ [1ex]
                 k_2^2-V_{22}-q_-^2 & & V_{21}
              \end{array}
               \right) \, . 
\end{equation}
It is easy
to check that the matrix ${\sf O}$ is orthogonal, ${\sf O} {\sf O}^{\rm T} =
{\sf O}^{\rm T} {\sf O} = {\sf I}$, and that it 
diagonalizes the matrix $[{\sf K}^2-{\sf V}]$,
\begin{equation}
  {\sf O} \left[ {\sf K}^2-{\sf V}\right] {\sf O}^{-1} = {\sf Q}^2 =
    \text{diag}(q_+^2,q_-^2) \, ,
    \label{diagonalization}
\end{equation}
where ${\sf Q}= \text{diag}(q_+,q_-)$ is the diagonal matrix whose
entries are $q_+$ and $q_-$.
 
We now define the column vector
\begin{equation}
  \widetilde{\sf \Psi} =\left\{
  \begin{array}{lc}
    {\sf O} {\sf \Psi} \, ,  & r<a \, ,  \\
    {\sf \Psi} \, ,  & r>a \, , 
    \end{array}\right.
     \label{tildevsnontilde}
\end{equation}
and combine it with Eq.~(\ref{diagonalization}) to convert
Eqs.~(\ref{cseqmforma})-(\ref{cseqmformb}) into a system of uncoupled
differential equations,
\begin{eqnarray}
  \widetilde{\sf \Psi}'' + {\sf Q}^2 \widetilde{\sf \Psi} =0\, ,
  \quad r<a \, ,
       \label{cseqmformwtildea}  \\
         \widetilde{\sf \Psi}'' +{\sf K}^2 \widetilde{\sf \Psi} =0 \, ,
         \quad r>a  \, .
     \label{cseqmformwtildeb}
\end{eqnarray}
We can now solve for $\widetilde{\sf \Psi}$, and then use
Eq.~(\ref{tildevsnontilde}) to obtain the original solution ${\sf \Psi}$.

The general solution of Eqs.~(\ref{cseqmformwtildea})
and~(\ref{cseqmformwtildeb}) can be written as
\begin{equation}
  \widetilde{\sf \Psi}(r;\vec{k}) = \left\{
  \begin{array}{lc}
   \cos({\sf Q}r) {\sf A} +  \sin ({\sf Q}r) {\sf B}\, ,  & r<a \, ,
            \\ [1ex]
          e^{-i{\sf K}r} {\sf C} + e^{i{\sf K}r} {\sf D} \, ,   &  r>a \,  , 
   \end{array}
  \right.
\end{equation}
where $\sin ({\sf Q}r)$, $\cos({\sf Q}r)$ and $e^{\pm i({\sf K}r)}$ are diagonal
matrices whose entries are, respectively, $\sin(q_{\pm}r)$, $\cos(q_{\pm}r)$
and $e^{\pm i k_{1,2}r}$, and where ${\sf A}$, ${\sf B}$, ${\sf C}$ and
${\sf D}$ are energy-dependent matrices or column vectors determined by boundary
conditions. By using Eq.~(\ref{tildevsnontilde}), the general solution
of Eqs.~(\ref{cseqmforma})-(\ref{cseqmformb}) can be written as
\begin{equation}
   {\sf \Psi}(r;\vec{k}) = \left\{
  \begin{array}{lc}
    {\sf O}^{-1} \Bigl(\cos({\sf Q}r)  {\sf A} + \sin ({\sf Q}r) {\sf B}
    \Bigr) \, ,  & r<a \, ,
            \\ [2ex]
          e^{-i{\sf K}r} {\sf C} + e^{i{\sf K}r} {\sf D} \, ,   &  r>a \,  . 
   \end{array}
  \right.
    \label{gensolpsi}
\end{equation}

\subsection{The regular solution, the Jost functions, and the $S$-matrix}

Let us denote by ${\sf \Phi}$ the $2\times 2$ matrix whose columns are
the regular solutions of Eqs.~(\ref{cseqmforma})
and~(\ref{cseqmformb}). The regular solution ${\sf \Phi}$ has the form of
Eq.~(\ref{gensolpsi}) and satisfies the boundary
conditions\footnote{The boundary conditions~(\ref{bcrs1})
and~(\ref{bcrs2}) are equivalent to requiring that
${\sf \Phi}(r;\vec{k}) \sim \sin ({\sf K}r)$ as $r\to 0$, which
is the way Refs.~\cite{TAYLOR,RAKITYANSKY} impose those boundary
conditions. Note however that, instead of
Eq.~(\ref{bcrs2}), Ref.~\cite{NEWTON} uses the boundary
condition ${\sf \Phi}'(0;\vec{k})={\sf I}$.}
\begin{eqnarray}
  & {\sf \Phi}(0;\vec{k})=0 \, , \label{bcrs1} \\
  & {\sf \Phi}'(0;\vec{k})={\sf K} \, , \label{bcrs2} \\
     & {\sf \Phi}(a+;\vec{k})= {\sf \Phi}(a-;\vec{k}) \, , \label{bcrs3}\\
      & {\sf \Phi}(a+;\vec{k})= {\sf \Phi}(a-;\vec{k}) \, . \label{bcrs4}
\end{eqnarray}
By imposing the boundary conditions~(\ref{bcrs1})-(\ref{bcrs4}) on the
general solution~(\ref{gensolpsi}), we obtain
\begin{equation}
    {\sf \Phi}(r;\vec{k}) = \left\{
  \begin{array}{lc}
    {\sf O}^{-1} \sin ({\sf Q}r) {\sf Q}^{-1} {\sf O} {\sf K} \, ,  & r<a \, ,
            \\ [2ex]
     \frac{i}{2}  \bigl( e^{-i{\sf K}r} {\sf F}_+ - e^{i{\sf K}r} {\sf F}_- \bigr)
             \, ,   &  r>a \,  , 
   \end{array}
  \right.
    \label{regsolphi}
\end{equation}
where the Jost matrices
${\sf F}_+ \equiv {\sf F}_+(E )\equiv {\sf F}_+(\vec{k})$ and
${\sf F}_-\equiv {\sf F}_-(E)\equiv {\sf F}_-(\vec{k})$ are related by
${\sf F}_-(\vec{k})={\sf F}_+(-\vec{k})$, and where the explicit expression
of ${\sf F}_+$ is determined by the boundary
conditions~(\ref{bcrs3}) and~(\ref{bcrs4}):
\begin{equation}
  {\sf F}_+(\vec{k})=
    e^{i{\sf K}a} \left( {\sf K}^{-1} {\sf O}^{-1} \cos ( {\sf Q}a)
  -i {\sf O}^{-1} {\sf Q}^{-1} \sin ({\sf Q}a)  \right) {\sf O} {\sf K} \, .
    \label{jostmatrix}
\end{equation}
The zeros of the determinant of the
Jost matrix coincide with the poles of the
$S$-matrix~\cite{NEWTON61,NEWTON,TAYLOR,RAKITYANSKY}, and
therefore determine
the bound and resonant energies. Because the determinants of
$e^{i{\sf K}a}$, ${\sf O}$ and ${\sf K}$ are all non-zero, the condition
\begin{equation}
      \text{det}\, {\sf F}_+(\vec{k})= 0 
    \label{zerodetera}
\end{equation}
is equivalent to 
\begin{equation}
      \text{det}\left( {\sf K}^{-1} {\sf O}^{-1} \cos ( {\sf Q}a)
  -i {\sf O}^{-1} {\sf Q}^{-1} \sin ({\sf Q}a)  \right) = 0 \, .
    \label{zerodeterb}
\end{equation}

\subsection{The Green function matrix}

The Jost solution ${\sf F}(r;\vec{k})$ satisfies
the following boundary conditions:
\begin{eqnarray}
  & {\sf F} (a+;\vec{k})= {\sf F} (a-;\vec{k})
           \, , \label{mcbcjs1}\\
           & {\sf F}'(a+;\vec{k})= {\sf F}'(a-;\vec{k})
              \, , \label{mcbcjs2} \\
              & {\sf F} (r;\vec{k}) \xrightarrow[r \to \infty]{\quad}
               e^{i{\sf K}r} \, .
     \label{mcbcjs3}
\end{eqnarray}
It is easy to show that
\begin{equation}
    {\sf F}(r;\vec{k}) =  \left\{
  \begin{array}{lc}
   {\sf O}^{-1} \left[  \cos ({\sf Q}(r-a)) {\sf O} +i {\sf Q}^{-1}
      \sin({\sf Q}(r-a))  {\sf O}{\sf K} \right] e^{i{\sf K}a}
        \, ,  & r<a \, ,
            \\ [1ex]
     e^{i{\sf K}r} \, ,   &  r>a \,  . 
   \end{array}
  \right.
     \label{mcjs}
\end{equation}
Clearly, ${\sf F}(0;\vec{k})={\sf K}^{-1} {\sf F}_+^{\rm T}(\vec{k}){\sf K}$. The
Green function can be written as~\cite{NEWTON61,MIZUYAMA}
\begin{equation}
  {\sf G}(r,s;E)=
  \left\{ \begin{array}{ll}
    \frac{2\mu}{\hbar ^2}
         {\sf F}(s;E)\left[{{\sf W}({\sf \Phi}, {\sf F})}\right]^{-1}
         {\sf \Phi}^{\rm T} (r;E) & r<s \, ,   \\ [2ex]
         \frac{2\mu}{\hbar ^2}{\sf \Phi}(s;E)
         \left[{{\sf W}^{\rm T}({\sf \Phi}, {\sf F})}\right]^{-1}
      {\sf F}^{\rm T}(r;E) & \quad r>s \, .
    \end{array}
  \right.
    \label{geforgfmcn}
\end{equation}
where the Wronskian is defined as
\begin{equation}
    {\sf W}({\sf \Phi}, {\sf F}) = {\sf \Phi}^{\rm T} {{\sf F}'}-
    [{\sf \Phi}']^{\rm T} {\sf F} \, .
\end{equation}
A straightforward calculation yields
\begin{equation}
  {\sf W}({\sf \Phi}, {\sf F}) =-{\sf F}_+^{\rm T}{\sf K} \, .
    \label{wronsmucg}
\end{equation}
By plugging Eq.~(\ref{wronsmucg}) into Eq.~(\ref{geforgfmcn}), and
restricting ourselves to the case $r>s$ for clarity,
we can write the Green function as
\begin{equation}
  {\sf G}(r,s;E)= -\frac{2\mu}{\hbar ^2}{\sf \Phi}(s;E)
  \frac{1}{{\sf F}_+(E){\sf K}}
       {\sf F}^{\rm T}(r;E) \, , \quad r>s \, .
       \label{greenfuninouno}
\end{equation}
The Green function is analytic everywhere except at the bound and resonant
energies.

\subsection{The bound and resonant states}

The bound and resonant states are column solutions
${\sf U}\equiv {\sf U}(r;E) \equiv {\sf U}(r;\vec{k})$ of
Eqs.~(\ref{cseqmforma}) and~(\ref{cseqmformb}) satisfying the
following boundary conditions:
\begin{eqnarray}
  & {\sf U}(0;\vec{k})=0 \, , \label{bcbrs1} \\
     & {\sf U}(a+;\vec{k})= {\sf U}(a-;\vec{k}) \, , \label{bcbrs2}\\
  & {\sf U}'(a+;\vec{k})= {\sf U}'(a-;\vec{k}) \, , \label{bcbrs3} \\
  & {\sf U}(r;\vec{k}) \xrightarrow[r\to \infty]{\quad}
      e^{i{\sf K}r} {\cal N} \, ,
     \label{bcbrs4}
\end{eqnarray}
where Eq.~(\ref{bcbrs4}) is the multichannel purely outgoing boundary
condition, and the column vector ${\cal N}$ contains the asymptotic
normalization constants.

By imposing the boundary conditions~(\ref{bcbrs1}) and~(\ref{bcbrs4}) on the
general solution~(\ref{gensolpsi}), we obtain:
\begin{equation}
  {\sf U} (r;\vec{k}) = \left\{
  \begin{array}{lc}
    {\sf O}^{-1}  \sin ({\sf Q}r) {\sf B} \, ,  & r<a \, ,
            \\ [1ex]
      e^{i{\sf K}r} {\cal N} \, ,   &  r>a \,  . 
   \end{array}
  \right.
     \label{usolbeac14}
\end{equation}
Imposing the boundary conditions~(\ref{bcbrs2}) and~(\ref{bcbrs3})
on the solution~(\ref{usolbeac14}) leads to
\begin{eqnarray}
  {\sf O}^{-1} \sin ({\sf Q}a) {\sf B} = e^{i{\sf K}a} {\cal N} \, , 
    \label{mcccond1}\\
 {\sf O}^{-1} {\sf Q} \cos ({\sf Q}a) {\sf B} = e^{i{\sf K}a} i{\sf K} {\cal N}
     \, .   \label{mcccond2}
\end{eqnarray}
This system of equations can be written in matrix form as
\begin{equation}
  \left( \begin{array}{cc}
    e^{i{\sf K}a} &  -{\sf O}^{-1} \sin ({\sf Q}a)  \\
    e^{i{\sf K}a} i{\sf K} & -{\sf O}^{-1} {\sf Q} \cos ({\sf Q}a)
  \end{array} \right) \left( \begin{array}{c}
    {\cal N} \\
    {\sf B}
  \end{array}\right) = 0 \, .
    \label{mafosyofcoeff}
\end{equation}
Equation~(\ref{mafosyofcoeff}) has a non-trivial solution only for energies
for which  the determinant of the matrix is zero:
\begin{equation}
  \text{det} \left( -e^{i{\sf K}a} {\sf O}^{-1} {\sf Q} \cos ({\sf Q}a)
  + e^{i{\sf K}a} i{\sf K}  {\sf O}^{-1} \sin ({\sf Q}a) \right) = 0 \, .
    \label{resonenpobc1}
\end{equation}
Because $e^{i{\sf K}a}$, ${\sf K}$ and ${\sf Q}$ have non-zero
determinant, we can write Eq.~(\ref{resonenpobc1}) as
\begin{equation}
  \text{det} \left( {\sf K}^{-1} {\sf O}^{-1} \cos ({\sf Q}a)
  - i  {\sf O}^{-1}{\sf Q}^{-1} \sin ({\sf Q}a) \right) =0 \, ,
    \label{resonenpobc2}
\end{equation}
which is just Eq.~(\ref{zerodeterb}). Hence, the purely outgoing
boundary condition leads to the same bound and resonant spectrum as the
condition that the determinant of the Jost matrix is zero.

By plugging Eq.~(\ref{mcccond1}) into Eq.~(\ref{usolbeac14}), we
obtain\footnote{Following
Ref.~\cite{RAKITYANSKY}, we are going to collectively
denote the bound and resonant energies (wave numbers) by $E_0$ ($k_0$).}
\begin{equation}
  {\sf U}(r;\vec{k}_{0}) = \left\{
  \begin{array}{lc}
    {\sf O}^{-1}  \sin ({\sf Q}_{0}r)[\sin({\sf Q}_{0}a)]^{-1} {\sf O}
      e^{i{\sf K}_{0}a}  {\cal N} \, ,  & r<a \, ,
            \\ [1ex]
      e^{i{\sf K}_{0}r} {\cal N} \, ,   &  r>a \,  , 
   \end{array}
  \right.
     \label{usolbeac15}
\end{equation}
where the column vector ${\cal N}$ is determined by normalization.

\subsection{Normalization of bound and resonant states}

For multichannel bound states and
resonances, there are two different normalizations in the literature,
Eq.~(41.7) of Ref.~\cite{BAZ} and Eq.~(12.92) of Ref.~\cite{RAKITYANSKY}. We
are going to normalize the
multichannel Gamow states by requiring that the residue of the Green
function at the resonant pole is factored out as the product of two Gamow
states, and show that such procedure yields the same normalization as
in Ref.~\cite{BAZ}.

By writing the Green function~(\ref{greenfuninouno}) as
\begin{equation}
  {\sf G}(r,s;E)= -\frac{2\mu}{\hbar ^2}{\sf \Phi}(s;E)
  [{\sf F}_-(E)]^{-1} [{\sf F}_-(E)] 
  \frac{1}{{\sf F}_+(E)}  \frac{1}{{\sf K}}
    {\sf F}^{\rm T}(r;E) \, , \quad r>s \, ,
\end{equation}
and by taking the residue at the pole $E_0$, we obtain
\begin{equation}
  \text{Res}\left[ {\sf G}(r,s;E)\right]_{E=E_{0}} =
      -\frac{2\mu}{\hbar ^2}
      {\sf \Phi}(s;E_{0})[{\sf F}_-(E_0)]^{-1} \,
        \text{Res}[{\sf S}_{\rm nu}(E)]_{E=E_0} \, 
      \frac{1}{{\sf K}_{0}}
           {\sf F}^{\rm T}(r;E_{0})\, , \quad r>s  \, ,
           \label{mcresgfun1}
\end{equation}
where the non-unitary $S$-matrix ${\sf S}_{\rm nu}(E)$ is defined as
\begin{equation}
     {\sf S}_{\rm nu}= {\sf F}_- \, {\sf F}_+^{-1} \, .
\end{equation}

We are now going to show that 
if the asymptotic normalization constants satisfy the following normalization
condition:
\begin{equation}
  {\cal N}{\cal N}^{\rm T} =
  i \frac{\mu}{\hbar ^2}
  \text{Res}[{\sf S}_{\rm nu}(E)]_{E=E_0} \, 
  \frac{1}{{\sf K}_{0}} \, ,
    \label{normamc}
\end{equation}
then the residue~(\ref{mcresgfun1}) of the Green function is factored out as
the product of two resonant states,
\begin{equation}
  \text{Res}\left[ {\sf G}(r,s;E)\right]_{E=E_{0}} =
       {\sf U}(s;E_{0}) {\sf U}^{\rm T}(r;E_{0}) \, , \quad r>s  \, .
       \label{facgfinfod}
\end{equation}
To show that Eq.~(\ref{facgfinfod}) holds when the Gamow states are
normalized according to Eq.~(\ref{normamc}), let us recall that 
on page~387 of Ref.~\cite{RAKITYANSKY}, it is shown that when the
regular and Jost solutions are analytically continued to the poles of
the $S$-matrix, they satisfy
\begin{eqnarray}
  && {\sf \Phi}(r;E_{0}) [{\sf F}_-(E_0)]^{-1} {\cal N} 
  = -\frac{i}{2} {\sf U} (r;E_{0}) \, , 
    \label{phivsuatpole} \\
  && {\sf F}(r;E_{0}) {\cal N} ={\sf U} (r;E_{0}) \, .  \label{fvsuatpole} 
\end{eqnarray}    
By plugging Eqs.~(\ref{normamc}), (\ref{phivsuatpole})
and~(\ref{fvsuatpole}) into Eq.~(\ref{facgfinfod}), we obtain
Eq.~(\ref{mcresgfun1}). Hence, the normalization condition~(\ref{normamc})
enables us to factorize the residue of the Green function as in
Eq.~(\ref{facgfinfod}).

The non-unitary $S$-matrix ${\sf S}_{\rm nu}(E)$ is related to the actual
unitary $S$-matrix ${\sf S}(E)$ by~\cite{TAYLOR,NEWTON,RAKITYANSKY}
\begin{equation}
  {\sf S}= {\sf K}^{1/2} {\sf S}_{\rm nu} {\sf K}^{-1/2}=
  {\sf K}^{1/2} {\sf F}_- \, {\sf F}_+^{-1}{\sf K}^{-1/2} \, .
     \label{usmatrix}
\end{equation}
Clearly, the poles of ${\sf S}(E)$  are the same as those of
${\sf S}_{\rm nu}(E)$. In terms of the (unitary) $S$-matrix, the
normalization condition~(\ref{normamc}) can be expressed as
\begin{equation}
  {\cal N}{\cal N}^{\rm T} =
                i \frac{\mu}{\hbar ^2}
 {\sf K}_{0}^{-1/2} \text{Res}[{\sf S}(E)]_{E=E_0} {\sf K}_{0}^{-1/2} \, .
  \label{mpanc}
\end{equation}
We can write Eq.~(\ref{mpanc}) in terms of the entries of those matrices as
\begin{equation}
  {\cal N}_{\alpha}{\cal N}_{\beta} =
                i \frac{\mu}{\hbar ^2}
 \frac{1}{\sqrt{k_{\alpha 0}k_{\beta0}}} \text{Res}[S_{\alpha \beta}(E)]_{E=E_0}  \, .
  \label{mpanc2}
\end{equation}
Equation~(\ref{mpanc2}) allows us to normalize the Gamow solutions of the
multichannel Schr\"odinger equation once we know the
$S$-matrix~(\ref{usmatrix}). Although we derived Eq.~(\ref{mpanc2})
for the simple case of square well potentials,
the result has quite general validity, as long as the reduced masses are the
same in all channels. For a bound state of energy $E_{\rm b}$,
Eq.~(\ref{mpanc2}) guarantees
that the wave function of a bound state is normalized to unity as follows:
\begin{equation}
  \int_0^{\infty} \left(  [u_{\rm bound, 1}(r;E_{\rm b})]^2 +
      [u_{\rm bound, 2}(r;E_{\rm b})]^2
    \right) dr =1 \, .
      \label{mulchnormbs}
\end{equation}
For resonant states, the integral in Eq.~(\ref{mulchnormbs}) needs to be
regularized.

We are now going to show that the normalization constants obtained
from Eq.~(\ref{mpanc2}) are the same as those of Ref.~\cite{BAZ} when
the state is bound. Following
Ref.~\cite{BAZ}, let us suppose that our multichannel system has a
bound state of energy
\begin{equation}
       E_0= \frac{\hbar^2 k_{10}^2}{2\mu} \, ,
\end{equation}
where $k_{10}$ is purely imaginary and we are assuming that $E_{\rm th,1}=0$. The
residue of the $S$-matrix as a function of $E$ is related to the
residue of the $S$-matrix as a function of
$k_1= \sqrt{ \frac{2\mu}{\hbar ^2} E}$ as follows: 
\begin{equation}
  \text{Res}[S_{\alpha \beta}(E)]_{E=E_0}= \frac{\hbar^2}{\mu}k_{10}\,
  \text{Res}[S_{\alpha \beta}(k_1)]_{k_1=k_{10}} \, .
  \label{relationbwus}
\end{equation}
Combining Eqs.~(\ref{mpanc2}) and~(\ref{relationbwus}) yields
\begin{equation}
  {\cal N}_{\alpha} \, {\cal N}_{\beta} =
                i 
                \sqrt{\frac{k_{10}^2}{k_{\alpha 0}k_{\beta 0}}} \, 
                \text{Res}[S_{\alpha \beta}(k_1)]_{k_1=k_{10}}  \, .
  \label{mpanc3}
\end{equation}
If we write this equation in terms of the velocities associated
with the wave vector $\vec{k}_0$, and we take into account that
${\cal N}_i$ can be chosen to be real for a bound state,
we can write Eq.~(\ref{mpanc3}) as
\begin{equation}
  {\cal N}_{\alpha} \, {\cal N}_{\beta}^* =
                i 
                \sqrt{\frac{v_{10}^2}{v_{\alpha 0}v_{\beta 0}}} \,
                \text{Res}[S_{\alpha \beta}(k_1)]_{k_1=k_{10}}  \, ,
  \label{mpanc5}
\end{equation}
which agrees with Eq.~(41.7) of Ref.~\cite{BAZ}.

\section{Numerical Calculations}
\setcounter{equation}{0}
\label{sec:calculation}

In this section, we present the numerical calculation of the bound and
resonant states, decay energy
spectra, partial decay constants, partial decay widths, and
branching fractions of the resonances of the single-channel
and two-channel square well potentials. In our calculations, we will use units
such that $a=1$ and $\hbar^2/2\mu =1$.

\subsection{The single-channel square well
  potential}

\begin{figure}[h!]
\begin{center}
              \epsfxsize=10cm
              \epsffile{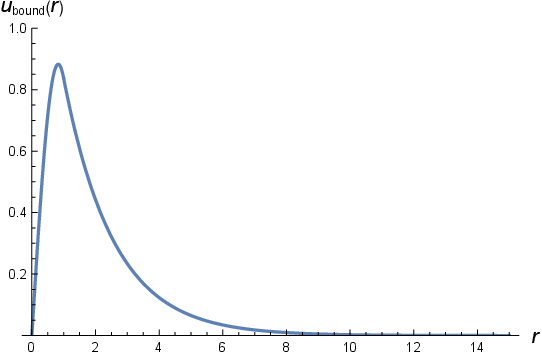}
\end{center}
\caption{
      Bound state wave function of energy $E_{\rm b}=-0.407101$.}
     \label{fig:sbswf}
\end{figure}

The bound and resonant states of the single-channel square well potential
are well known. For the potential depth $V_0=-4$, there is a bound
state of energy $E_{\rm b}=-0.407101$. The corresponding bound-state wave
function, Eq.~(\ref{spusolbeac15}), is shown in Fig.~\ref{fig:sbswf}.

\noindent The norm of the bound state in Fig.~\ref{fig:sbswf} is unity,
thereby confirming that Zeldovich's normalization, Eq.~(\ref{spanc}), ensures
that bound states are normalized to unity.

\begin{figure}[h!]
\begin{center}
              \epsfxsize=10cm
              \epsffile{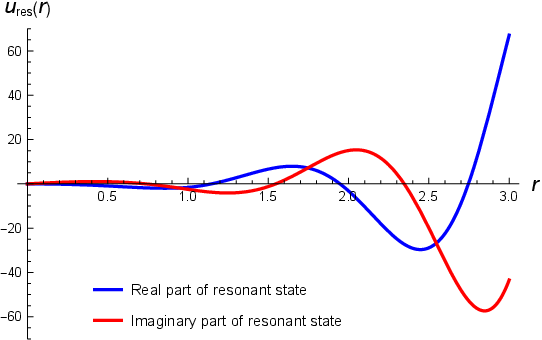}
\end{center}
\caption{
  Real and imaginary parts of the Gamow eigenfunction of the resonance of energy
  $E_{\rm res}=12.7131- i\, 12.9422$.}
     \label{fig:srswf}
\end{figure}

The square well potential has an infinite number of resonant
energies. When $V_0=-4$, the resonant energy that is closest to the threshold
is $E_{\rm res}=12.7131- i\, 12.9422$. The real and imaginary parts of the
associated Gamow eigenfunction, Eq.~(\ref{spusolbeac15}), are plotted
in Fig.~\ref{fig:srswf}.
As is typical of resonant eigenfunctions, the real and imaginary parts
have small oscillations in the potential region and start oscillating wildly
as you move away from the potential region.

\begin{figure}[h!]
\begin{center}
              \epsfxsize=10cm
              \epsffile{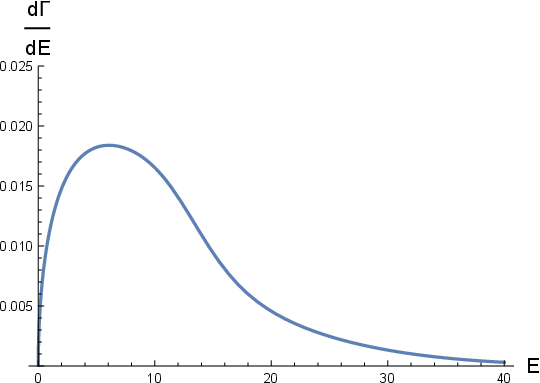}
\end{center}
\caption{
  Decay energy spectrum of the resonance of energy
  $E_{\rm res}=12.7131- i\, 12.9422$.}
     \label{fig:srsdd}
\end{figure}

The decay energy spectrum of this resonance, Eq.~(\ref{inteGam3}), is
plotted in Fig.~\ref{fig:srsdd}.
Because the resonance is broad, its decay energy spectrum is a fairly
broad, distorted Lorentzian. The decay constant and width of this
resonance are $\Gamma = 0.296381$ and $\overline{\Gamma} = 7.67165$.

\subsection{The two coupled-channel square well potentials}

\begin{figure}[h!]
\begin{center}
              \epsfxsize=10cm
              \epsffile{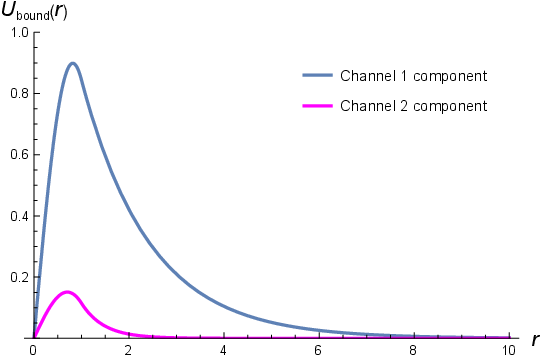}
\end{center}
\caption{
  Channel-1 and -2 components of the bound state of energy
  $E_{\rm b}=-0.486193$.}
     \label{fig:cbswf}
\end{figure}

For the two-channel square potential, we will use the potential depths
$V_{11}=V_{22}=-4$, and the threshold energies
$E_{{\rm th},1}=0$ and $E_{{\rm th},2}=4$. When
there is no coupling ($V_{12}=0$), each channel has one bound state
with energies $E_{\rm b}^{(0)}=-0.407101$ (channel 1) and
$E_{\rm b}^{(0)}=3.592899$ (channel~2). When the coupling is turned
on and set to $V_{12}=-1$, the bound state of channel~1 remains bound and
its energy becomes $E_{\rm b}=-0.486193$, whereas the bound state of channel~2
becomes a resonance of energy
$E_{\rm res}=3.66303-i\,0.0581467$~\cite{HAN,GROZDANOV}. It should be
noted that, when the coupling between the channels is turned on, the resonance
that originated from a bound state of channel~2 can decay into both channel~1
and channel~2.
The channel components of the
bound state wave function, Eq.~(\ref{usolbeac15}), are shown in
Fig.~\ref{fig:cbswf}.
When we calculate the norm of the bound state using Eq.~(\ref{mulchnormbs}),
we indeed obtain unity, thereby confirming that Eq.~(\ref{mpanc2}) is
the correct normalization condition of multichannel bound and resonant states.

For the two-channel squared well potential, the Riemann surface has four
interconnected sheets~\cite{RAKITYANSKY}. The bound state $E_{\rm b}=-0.486193$
lies in the physical sheet, characterized by ${\rm Im}(k_1)>0$,
${\rm Im}(k_2)>0$. The resonance $E_{\rm res}=3.66303-i\,0.0581467$ lies
in the sheet characterized by ${\rm Im}(k_1)<0$, ${\rm Im}(k_2)>0$. Therefore,
to calculate numerically the bound and resonant energies
as zeros of the Jost determinant~(\ref{zerodetera}), it is necessary to set
the signs of the imaginary parts of the channel wave numbers properly,
otherwise the corresponding zeros may not correspond to actual bound and
resonant energies.

\begin{figure}[h!]
\begin{center}
              \epsfxsize=10cm
              \epsffile{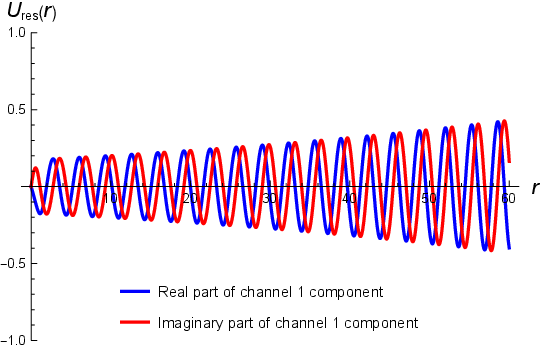}
\end{center}
\caption{
  Real and imaginary parts of the channel-1 component of the
  Gamow state of energy $E_{\rm res}=3.66303-i\,0.0581467$.}
     \label{fig:crswf1}
\end{figure}

\begin{figure}[h!]
\begin{center}
              \epsfxsize=10cm
              \epsffile{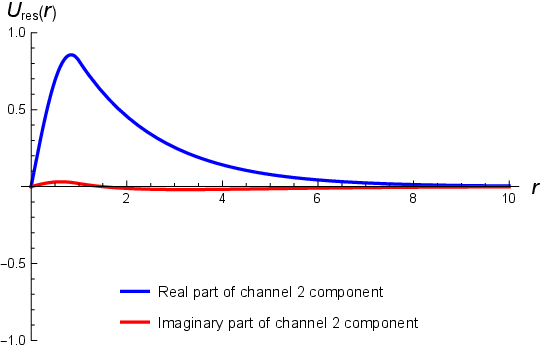}
\end{center}
\caption{
  Real and imaginary parts of the channel-2 component of the
  Gamow state of energy $E_{\rm res}=3.66303-i\,0.0581467$.}
     \label{fig:crswf2}
\end{figure}

The real and imaginary parts of the channel components of the resonant state,
Eq.~(\ref{usolbeac15}),
of energy $E_{\rm res}=3.66303-i\,0.0581467$ are plotted in
Figs.~\ref{fig:crswf1} and~\ref{fig:crswf2}. The channel-1 component
is very similar to a single-channel resonance: oscillatory in the region
of the potential, and with exponentially increasing oscillations as you move
away from the potential region.
Since this resonance is right below the threshold of channel~2, its
channel-2 component looks very similar to a single-channel bound state,
albeit with a non-zero imaginary part.

\begin{figure}[h!]
\begin{center}
              \epsfxsize=10cm
              \epsffile{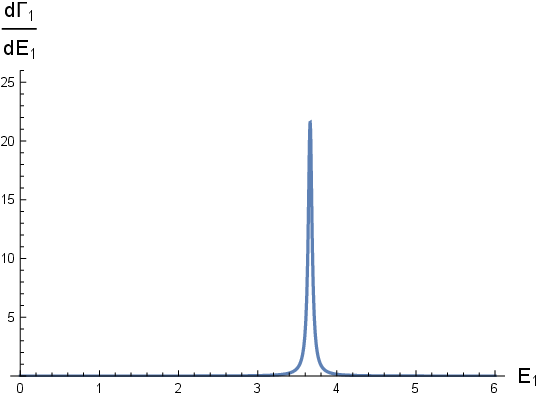}
\end{center}
\caption{Channel-1 decay energy spectrum of the resonance
   $E_{\rm res}=3.66303-i\,0.0581467$.}
     \label{fig:cdd1}
\end{figure}

As mentioned above, this resonance can decay both into channel~1 and~2. The
decay energy spectrum for decay into channel~1, Eq.~(\ref{lsmulticg91}),
is shown in Fig.~\ref{fig:cdd1}.
The channel-1 decay energy spectrum is given by a sharp Lorentzian-like peak
centered around the real part of the resonant energy.

\begin{figure}[ht!]
\begin{center}
              \epsfxsize=10cm
              \epsffile{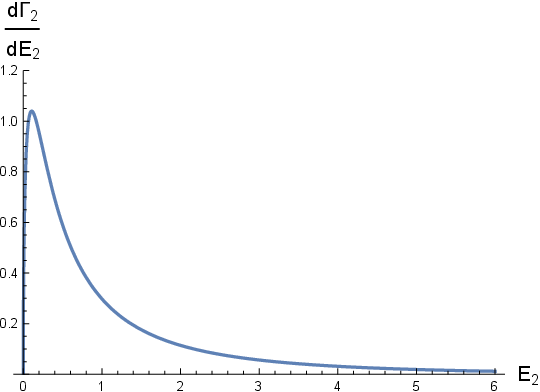}
\end{center}
\caption{Channel-2 decay energy spectrum of the resonance
   $E_{\rm res}=3.66303-i\,0.0581467$.}
     \label{fig:cdd2}
\end{figure}

The decay energy spectrum into channel~2, Eq.~(\ref{lsmulticg92}), is shown in 
Fig.~\ref{fig:cdd2}.
The asymmetric ``half-peak'' structure of Fig.~\ref{fig:cdd2} is typical of
resonances right below the threshold of a channel. It should be noted that
the decay spectrum
of Fig.~\ref{fig:cdd2} is expressed in terms of the kinetic energy
$E_2$ of channel 2 instead of the total energy $E$, and
therefore the threshold energy $E=E_{{\rm th},2}=4$ corresponds to
$E_2= E-E_{{\rm th},2}=0$.

By integrating the differential partial decay constants of
Eqs.~(\ref{lsmulticg91}) and~(\ref{lsmulticg92}) over
the positive real line, we can obtain
the partial decay constants:
\begin{equation}
       \Gamma _1 = 1.96087 \, , \quad \Gamma_2 = 1.00172 \, .
\end{equation}
The partial decay widths and branching fractions are given by
Eq.~(\ref{bflth}),
\begin{eqnarray}
  \overline{\Gamma}_1 = 0.228036 \, , \quad
  & \overline{\Gamma}_2 = 0.116493 \, , \\
  {\cal B}_1 = 66.19 \% \, , \quad
    & {\cal B}_2 = 33.81 \% \, .
\end{eqnarray}
Hence, this resonance has about a $66\%$ chance to decay into channel~1, and
about a $34\%$ chance to decay into channel~2.

\section{Conclusions}
\setcounter{equation}{0}
\label{sec:conclusions}

We have extended the Gamow Golden Rule to resonances that
have several decay modes. The resulting multichannel Gamow
Golden Rule is formally the same as the single-channel one: The decay spectrum
of a given channel is given by the Lorentzian multiplied by the matrix element
of the interaction, although the matrix element involves the interference
of all the channels to which the resonance can decay.

We have obtained the multichannel Gamow Golden Rule in the coupled-channel
approximation. We have also introduced a normalization of the multichannel
Gamow states that ({\it i}) is a natural generalization of the original
single-channel normalization proposed by Zeldovich, and ({\it ii}) coincides
with the normalization of Ref.~\cite{BAZ}.

We have exemplified the multichannel Gamow Golden Rule using two
coupled-channel square well potentials. We have calculated numerically
the partial
(channel) decay distributions, the partial decay constants, the partial
decay widths, and the branching fractions. We have also
checked numerically that a multichannel bound state is normalized to unity,
thereby confirming that the normalization of multichannel bound and resonant
states is correct.

To describe resonances, established nonrelativistic scattering
theory~\cite{NEWTON,BAZ,TAYLOR,MAREK,RAKITYANSKY} focuses almost
exclusively on cross sections. The
Gamow Golden Rule complements scattering theory by describing
decay (invariant mass) distributions, a description that is lacking in the
nuclear physics literature.

The Gamow Golden Rule is useful only in situations where we can
disentangle the resonant signal from the background and we can neglect
their interference. The Gamow Golden Rule is therefore
a ``pole-only'' description of decay distributions. Hence, the Gamow Golden
Rule does not describe situations where the interference of the resonance
and the background is important (e.g., Fano resonances~\cite{FANO}). 

This pole-only description of decay distributions is conceptually similar to
the truncated, pole-only resonant expansions. Wave
functions can be expanded in terms of the Gamow states
and continuum background states (see for example review~\cite{CJP}). When
the wave function can be well approximated by the Gamow state, and therefore
the background and its interference with the resonance can be neglected,
one obtains a truncated, pole-only resonant expansion.

Phenomenological, non-Hermitian Hamiltonians have been used extensively
in nuclear physics, as for example in the random-matrix formulation of
statistical spectroscopy~\cite{SOKOLOV}. Truncated, pole-only resonant
expansions can be used to justify such phenomenological non-Hermitian
Hamiltonians~\cite{CJP}. However, it is an open question if the Gamow
Golden Rule can be related to the statistical theory of spectra~\cite{SOKOLOV}.

Thresholds have an effect on the decay distributions of the Gamow Golden Rule,
even for single-channel resonances, where the presence of a threshold
close to the resonance distorts the lower end of the decay
distribution (see Fig.~\ref{fig:srsdd} and Ref.~\cite{NPA17}). In the
two-channel example we chose, the effect of
the threshold is to produce a half-peak distribution for the decay into the
second channel, which is also the typical shape of the decay distribution of
single-channel virtual states~\cite{NPA17}. However, it is an open
question to describe the exact effect of thresholds on decay distributions. In
particular, it is an open question if the decay distributions of
multichannel resonances exhibit features similar to Wigner
cusps~\cite{WIGNER}.

The decay distributions of the Gamow Golden Rule are in general
asymmetric, see Fig.~\ref{fig:srsdd}. Even single-channel
resonances that are isolated but broad have decay distributions that
are asymmetric and exhibit small oscillations away from the
central resonant peak~\cite{NPA17}. Typically, only resonances that are
sharp, isolated and far from thresholds have symmetric, quasi-Lorentzian
lineshapes. 

Besides describing decay distributions, the usefulness of the Gamow
Golden Rule hinges on being able to account for the decay rate $\gr$ that
arises from the imaginary part of the pole of the $S$-matrix,
$E_{\rm res} = \er - i \gr /2$. It can be shown that, when the absolute value
squared of the Gamow state in the momentum representation is normalized
to unity, the decay width of the Gamow Golden Rule is exactly the
same as $\gr$, no matter how close the resonance is to the a threshold,
or how strong or weak the interaction is. This result will be presented
elsewhere.

Altogether, we have established the theoretical and conceptual foundations
of the multichannel Gamow Golden Rule. The results can now be incorporated
into more realistic nuclear physics models, such as the Gamow Shell
Model~\cite{MAREK}, to describe experimental decay
distributions. Further work in this direction is in progress.

\section*{Acknowledgment}

The authors thank Sergei Rakityansky for enlightening correspondence. The
work of RIB~has been partially supported by CONICET (Consejo Nacional de
Investigaciones Cient\'ificas y T\'ecnicas), under the Grant
No.~80020210200068UR, and Universidad Nacional de Rosario,
Grant No.~PIP-0930, Argentina.

\appendix
\section{The coupled-channel approximation of the Gamow Golden Rule in
  the angular momentum basis}
\label{app.a}

In this Appendix, we are going to show that when the potentials are
spherically symmetric, the Golden Rule can be expressed as in
Eq.~(\ref{lsmulticg5ss}).

Let us first recall the expansion of plane waves in terms of spherical
harmonics~\cite{TAYLOR}:
\begin{equation}
  \phi _{\alpha} ^{\rm free}(\vec{x})=
    \frac{1}{(2\pi\hbar)^{3/2}} e^{i\vec{p}_{\alpha}\cdot \vec{x}/\hbar} =
          \sum_{l',m'} \sqrt{\frac{1}{\mu_{\alpha} p_{\alpha}}}
             \frac{\psi_{l'}^0(r;E_{p_{\alpha}})}{r} 
          Y_{l'}^{m'}(\hat{x}) Y_{l'}^{m'}(\hat{p}_{\alpha})^*  \, ,
\end{equation}
where $\psi_{l'}^0(r;E_{p_{\alpha}})$ is given in Eq.~(\ref{angulmombdefree}). The
matrix element of the interaction becomes
\begin{eqnarray}
   \langle\phi _{\alpha} ^{\rm free}|\bar{V}_{\alpha,\beta}| 
         \phi ^{\rm res}_{\beta}\rangle &=& 
         \int d^3x \,  \frac{1}{(2\pi\hbar)^{3/2}}
             e^{-i\vec{p}_{\alpha}\cdot \vec{x}/\hbar}
             \bar{V}_{\alpha,\beta}(\vec{x})
             \frac{u_{\beta,l}(r;E_{\rm res})}{r}Y_l^m(\hat{x} ) \nonumber    \\
           &=& 
          \int r^2 dr d\Omega_{\hat{x}} \, 
 \sum_{l',m'} \sqrt{\frac{1}{\mu_{\alpha} p_{\alpha}}}
             \frac{\psi_{l'}^0(r;E_{p_{\alpha}})}{r} 
          Y_{l'}^{m'}(\hat{x})^* Y_{l'}^{m'}(\hat{p}_{\alpha})
 \bar{V}_{\alpha,\beta}(r)
\frac{u_{\beta,l}(r;E_{\rm res})}{r}Y_l^m(\hat{x} ) \nonumber \\
   &=& 
 \sqrt{\frac{1}{\mu_{\alpha} p_{\alpha}}} \sum_{l',m'}  \int dr  \, 
             \psi_{l'}^0(r;E_{p_{\alpha}})          
 \bar{V}_{\alpha,\beta}(r)
 u_{\beta,l}(r;E_{\rm res})  Y_{l'}^{m'}(\hat{p}_{\alpha})
 \int d\Omega_{\hat{x}} Y_{l'}^{m'}(\hat{x})^*  Y_l^m(\hat{x})   \nonumber \\
 &=& 
 \sqrt{\frac{1}{\mu_{\alpha} p_{\alpha}}}  \int dr  \, 
             \psi_{l}^0(r;E_{p_{\alpha}})          
 \bar{V}_{\alpha,\beta}(r)
 u_{\beta,l}(r;E_{\rm res})  Y_{l}^{m}(\hat{p}_{\alpha})  \nonumber \\ 
     &=&  \sqrt{\frac{1}{\mu_{\alpha} p_{\alpha}}}
       \langle E_{p_{\alpha}}|\bar{V}_{\alpha,\beta}|E_{\rm res}, \beta \rangle _l \,
         Y_{l}^{m}(\hat{p}_{\alpha}) \, .
    \label{maelpte}
\end{eqnarray}
By substituting Eq.~(\ref{maelpte}) into Eq.~(\ref{lsmulticg5}), and
by using the fact that
$d^3p_\alpha= dE_{p_{\alpha}} \,
\mu_{\alpha} \sqrt{2\mu_{\alpha} E_{p_{\alpha}}} \, d\Omega_{{\hat p}_{\alpha}}$, we
obtain that
\begin{equation}
      \frac{d\Gamma_{\alpha}}{dE_{p_\alpha}} =
      \int \mu_{\alpha} \sqrt{2\mu_{\alpha} E_{p_{\alpha}}} \, d\Omega_{{\hat p}_{\alpha}}
       \frac{1}{(E - \er)^2+ (\gr /2)^2} 
          \left| \sum _{\beta =1}^N
\sqrt{\frac{1}{\mu_{\alpha} p_{\alpha}}}
       \langle E_{p_{\alpha}}|\bar{V}_{\alpha,\beta}|E_{\rm res}, \beta \rangle _l \,
       Y_{l}^{m}(\hat{p}_{\alpha})
                 \right| ^2 \, ,   
\end{equation}
which can be simplified to Eq.~(\ref{lsmulticg5ss}).

\section{A second derivation of the coupled-channel approximation of the
Gamow Golden Rule
  in the angular-momentum basis}
\label{app.b}

In this Appendix, we are derive Eqs.~(\ref{lsmulticg91})
and~(\ref{lsmulticg92}) using the explicit
expressions of the Schr\"odinger equations of the coupled-channel system for
spherically symmetric potentials. For the sake of simplicity,
we will restrict ourselves to spinless particles of
zero angular momentum. The $s$-wave solution ${\sf \Psi}^+$ of
Eqs.~(\ref{channel1a})
and~(\ref{channel2a}) that has scattering boundary conditions can be written
as follows~\cite{NEWTON61}:
\begin{equation}
  {\sf \Psi}^+(r;\vec{p}) = {\sf \Psi}^0(r;\vec{p}) +
  \int_0^{\infty} ds \, {\sf G}_0(r,s;\vec{p}) \bar{\sf V}(s)
      {\sf \Psi} ^+ (s;\vec{p}) \, ,
       \label{LSswave2channel}
\end{equation}
where ${\sf \Psi}^0(r;\vec{p})$ is the matrix whose columns are the two
linearly independent solutions in Eq.~(\ref{freeparsoechannel}) for
$l=0$,
${\sf G}_0(r,s;\vec{p})$ is the free Green function matrix, and
$\bar{\sf V}$ is the $2\times 2$ matrix whose entries are
the potentials $\bar{V}_{ij}$. It is convenient to write
Eq.~(\ref{LSswave2channel}) in Dirac's bra-ket notation as
\begin{equation}
 \ket{{\sf \Psi} ^+} = \ket{{\sf \Psi}^0} +
  \frac{1}{E-H_0+i0} \bar{\sf V} \ket{{\sf \Psi} ^+}  .
\end{equation}
The integral equation for bound and resonant states is the same, but without
the free incoming part:
\begin{equation}
 \ket{{\sf \Psi}_{\rm res}} = 
 \frac{1}{E_{\rm res}-H_0+{\rm i}0} \bar{\sf V} \ket{{\sf \Psi}_{\rm res}}  .
    \label{integraleqnrad}
\end{equation}
By taking the inner product of Eq.~(\ref{integraleqnrad}) with the
free-particle wave functions of Eq.~(\ref{freeparsoechannel}), we get
\begin{equation}
     \braket{E_{p_{i}}|{\sf \Psi}_{\rm res}} = \frac{1}{E_{\rm res}-E} \,     
    \langle E_{p_{i}}|\bar{\sf V}|{\sf \Psi}_{\rm res}\rangle \, , 
\end{equation}
which can be written explicitly for each channel as
\begin{eqnarray}
   \langle E_{p_1}|{\sf \Psi}_{\rm res}\rangle = \frac{1}{E_{\rm res}-E}
     \left(
    \langle E_{p_1}|\bar{V}_{11}|E_{\rm res},1 \rangle
   + \langle E_{p_1}|\bar{V}_{12}|E_{\rm res} ,2 \rangle \right) \, ,  \\
    \langle E_{p_2}|{\sf \Psi}_{\rm res}\rangle = \frac{1}{E_{\rm res}-E}
     \left(
   \langle E_{p_2}|\bar{V}_{21}|E_{\rm res} ,1 \rangle
   + \langle E_{p_2}|\bar{V}_{22}|E_{\rm res} ,2 \rangle \right) \, . 
\end{eqnarray}
By taking the absolute value squared of these equations, we obtain the
partial differential decay constants, which coincide with 
those in Eqs.~(\ref{lsmulticg91}) and~(\ref{lsmulticg92}) when $l=0$.

\section{The Single-Channel Square Well Potential}
\setcounter{equation}{0}
\label{app.sphesymm-single}

\subsection{The Schr\"odinger equation and its general solution}

The radial Schr\"odinger equation for a particle of mass $m$ and
zero angular momentum is
\begin{equation}
      -\frac{\hbar ^2}{2m} \frac{d^2\psi}{dr^2} + \bar{V}(r) \psi (r)
      = E \psi (r) \, .
       \label{spscheq1}
\end{equation}
By defining $k=\sqrt{\frac{2m}{\hbar ^2}E}$ and
$V(r)= \frac{2m}{\hbar ^2}\bar{V}(r)$, Eq.~(\ref{spscheq1}) becomes
\begin{equation}
      - \frac{d^2\psi}{dr^2} + V(r) \psi (r)
      = k^2\psi (r) \, .
         \label{spscheq3}
\end{equation}
For the square barrier/well potential of height/depth $V_0$ and range $a$,
\begin{equation}
  V (r) =
  \left\{ \begin{array}{cc}
    V_{0}  &  r<a \, ,  \\
    0      & r>a \, ,
  \end{array}
    \right.  
\end{equation}
Eq.~(\ref{spscheq3}) can be written as
\begin{eqnarray}
  \psi '' + q^2 \psi =0 \, , &  & r<a \, ,
  \label{spcseqmforma} \\
   \psi '' + k^2 \psi =0 \, , & & r>a \, ,
     \label{spcseqmformb}
\end{eqnarray}
where $q^2 = k^2-V_0$. The general solution of
Eqs.~(\ref{spcseqmforma})-(\ref{spcseqmformb}) can be written as
\begin{equation}
  \psi (r;k) = \left\{
  \begin{array}{lc}
    a(k) \cos (qr) +b(k) \sin (qr) \, ,  & r<a \, ,
            \\ [1ex]
        c(k) \, e^{-ikr} +d(k)\, e^{ikr}  \, ,   &  r>a \,  , 
   \end{array}
  \right.
      \label{spgensolpsi}
\end{equation}
where $a(k)$, $b(k)$, $c(k)$ and $d(k)$ are functions of the wave number
(energy) determined by boundary conditions.

\subsection{The regular solution, the Jost functions, and the $S$-matrix}

The regular solution $\phi(r;k)$ satisfies the following boundary
conditions:
\begin{eqnarray}
  & \phi (0;k)=0 \, , \label{spbcrs1} \\
  & \phi'(0; k)= k \, , \label{spbcrs2} \\
   & \phi (a+;k)= \phi (a-;k) \, , \label{spbcrs3}\\
      & \phi(a+;k)= \phi(a-;k) \, . \label{spbcrs4}
\end{eqnarray}
By imposing the boundary conditions~(\ref{spbcrs1})-(\ref{spbcrs4}) on the
general solution~(\ref{spgensolpsi}), we obtain
\begin{equation}
    \phi (r;k) = \left\{
  \begin{array}{lc}
    \frac{k}{q} \sin (qr) \, ,  & r<a \, ,
            \\ [2ex]
     \frac{i}{2}  \bigl(f_+(k) e^{-ikr} - f_-(k) e^{ikr} \bigr)
             \, ,   &  r>a \,  , 
   \end{array}
  \right.
    \label{spregsolphi}
\end{equation}
where the Jost functions $f_{\pm}(k)$ are given by
\begin{eqnarray}
  && f{_+}(k)= e^{ika} \left( \cos(qa) -i\frac{k}{q} \sin (qa)\right) \, ,
   \label{spjfunc} \\
  && f{_-}(k)=f{_+}(-k) \, . 
\end{eqnarray}
The $S$-matrix is given by
\begin{equation}
  S(k)=\frac{f{_-}(k)}{f{_+}(k)} \, .
  \label{spsmatrix}
\end{equation}

\subsection{The Green function}

In order to obtain the Green function, let us
first obtain the Jost solution $f(r;k)$, which satisfies the
following boundary conditions:
\begin{eqnarray}
     & f(a+;k)= f(a-;k) \, , \label{spbcjs1}\\
  & {f}'(a+;k)= {f}'(a-;k) \, , \label{spbcjs2} \\
  & f(r;k) \xrightarrow[r \to \infty]{\quad}  e^{ikr} \, ,
     \label{spbcjs3}
\end{eqnarray}
It is easy to show that
\begin{equation}
    f(r;k) =  \left\{
  \begin{array}{lc}
    e^{ika} \left[ \cos q(r-a)+i\frac{k}{q} \sin q(r-a)\right]   \, ,  & r<a \, ,
            \\ [1ex]
     e^{ikr} \, ,   &  r>a \,  . 
   \end{array}
  \right.
     \label{spjs}
\end{equation}
Clearly, $f(0;k)=f_+(k)$. In addition,
\begin{equation}
  \phi (r;k) = \frac{i}{2} \left[ f_+(k)f(r;-k) -f_-(k) f(r;k)\right] \, .
\end{equation}

As is well known, the Green function can be written as
\begin{equation}
  G(r,s;E)= \frac{2m}{\hbar ^2}\frac{\phi(r_<;k)f(r_>;k)}{W(\phi, f)} \, ,
\end{equation}
where $W(\phi, f)$ is the Wronskian of $\phi(r;k)$ and $f(r;k)$, and
where $r_<$ ($r_>$) is the smallest (largest) of $r$ and $s$. Since
$W(\phi, f)=-kf_+(k)$, we can write the Green function as
\begin{equation}
  G(r,s;E)= -\frac{2m}{\hbar ^2} \frac{\phi(r_<;k)f(r_>;k)}{kf_+(k)} \, .
  \label{greenfunction}
\end{equation}

\subsection{The bound and resonant states}

The bound and resonant states $u(r;k)$ are solutions of
Eqs.~(\ref{spcseqmforma}) and~(\ref{spcseqmformb}) satisfying the
following boundary conditions:
\begin{eqnarray}
  & u(0;k)=0 \, , \label{spbcbrs1} \\
     & u(a+;k)= u(a-;k) \, , \label{spbcbrs2}\\
  & u'(a+;k)= u'(a-;k) \, , \label{spbcbrs3} \\
  & u(r;k) \xrightarrow[r \to \infty]{\quad} {\cal N}  e^{ikr} \, ,
     \label{spbcbrs4}
\end{eqnarray}
where Eq.~(\ref{spbcbrs4}) is the purely outgoing boundary
condition, and ${\cal N}$ is the asymptotic
normalization constant.

By imposing the boundary conditions~(\ref{spbcbrs1}) and~(\ref{spbcbrs4}) on
the general solution~(\ref{spgensolpsi}), we obtain:
\begin{equation}
  u(r;k) = \left\{
  \begin{array}{lc}
    b(k)  \sin (qr)  \, ,  & r<a \, ,
            \\ [1ex]
     {\cal N}  e^{ikr} \, ,   &  r>a \,  . 
   \end{array}
  \right.
     \label{spusolbeac14}
\end{equation}
Imposing the boundary conditions~(\ref{spbcbrs2}) and~(\ref{spbcbrs3})
on the solution~(\ref{spusolbeac14}) leads to
\begin{eqnarray}
  b (k) \sin (q a) = {\cal N} e^{ik a} \, ,  \label{spcontatr0}  \\
  b(k) \, q \cos (q a) = ik {\cal N} e^{ik a} \, .
\end{eqnarray}
This system can be written in matrix form as
\begin{equation}
  \left( \begin{array}{cc}
    e^{ika} &  - \sin (qa)  \\
    ik e^{ika} & -q \cos (qa)
  \end{array} \right) \left( \begin{array}{c}
    {\cal N} \\
    b(k)
  \end{array}\right) = 0  \, .
     \label{matrixformrescond}
\end{equation}
Equation~(\ref{matrixformrescond}) has a non-trivial solution only for
energies that make the
determinant of the matrix zero:
\begin{equation}
  e^{ik_{0}a} \left[ q_{0} \cos (q_{0}a) -ik_{0}
    \sin (q_{0}a) \right] =0 \, ,
    \label{spresonenpobc1}
\end{equation}
which is equivalent to the condition that the Jost function~(\ref{spjfunc})
is zero. Hence, the purely outgoing
boundary condition leads to the same bound and resonant spectrum as the
condition that the Jost function is zero, which is equivalent
to the condition that bound and resonant energies are poles of the
$S$-matrix~(\ref{spsmatrix}).

By plugging Eq.~(\ref{spcontatr0}) into the solution~(\ref{spusolbeac14}),
we obtain
\begin{equation}
  u(r;k_{0}) =  \left\{
  \begin{array}{lc}
    {\cal N}\frac{e^{ik_{0}a}}{\sin(q_{0}a)}
        \sin (q_{0}r)  \, ,  & r<a \, ,
            \\ [1ex]
     {\cal N}  e^{ik_{0}r} \, ,   &  r>a \,  . 
   \end{array}
  \right.
     \label{spusolbeac15}
\end{equation}
where ${\cal N}$ is determined by  normalization.

\subsection{Normalization of bound and resonant states}

If we normalize (regularize) the natural square of the bound (resonant)
eigenfunction,
\begin{equation}
  \int_0^{\infty} dr \, [u(r;k_{0})]^2 = 1 \, ,
\end{equation}
then it is well known that ${\cal N}$ is given in terms of the residue
of the $S$-matrix as
\begin{equation}
  {\cal N}^2 = \frac{im}{\hbar ^2 k_{0}} \text{Res}[S(E)]_{E=E_{0}}
      \, ,
    \label{spanc}
\end{equation}
where the residue is obtained by using $E$ (not $k$) as the
independent variable.

There are three main procedures to obtain the normalization
condition~(\ref{spanc}): Zeldovich's regulator, complex
scaling, and factorization of the residue of the Green function at a
resonant energy as the product of two Gamow eigenfunctions. The last one is the
procedure we are going to use.

The Green function~(\ref{greenfunction}) is analytic everywhere except at
the zeros of the denominator, which coincide with the bound and resonant
energies. The residue at the poles can be written as
\begin{equation}
  \text{Res}\left[ G(r,s;E)\right]_{E=E_{0}} =
  -\frac{2m}{\hbar ^2}\frac{\phi(r_<;E_{0})f(r_>;E_{0})}{k_{0}
    \dot{f}_+(E_{0})} \, ,
     \label{spresgf1}
\end{equation}
where $\dot{f}_+(E_{0})$ is the derivative of ${f}_+(E)$ with respect to $E$
evaluated at $E_0$. Because
$\phi(r;E_{0}) = -\frac{i}{2} f_-(k_{0})\frac{1}{ {\cal N}} u(r;E_{0})$ 
and ${\cal N} f(r;E_{0}) = u(r;E_{0})$, we can write Eq.~(\ref{spresgf1}) as
\begin{equation}
  \text{Res}\left[ G(r,s;E)\right]_{E=E_{0}} =
  u(r_<;E_{\rm res})\frac{1}{{\cal N}^2}\frac{2m}{\hbar ^2}
      \frac{i f_-(E_{0})}{2k_{0}\dot{f}_+(E_{0})}
      u(r_>;E_{\rm res})  \, .
\end{equation}
If we require that this residue be factored out as the product of two
Gamow eigenfunctions,
\begin{equation}
  \text{Res}\left[ G(r,s;E)\right]_{E=E_{0}} =
  u(r_<;E_{0}) u(r_>;E_{0})  \, ,
\end{equation}
then the asymptotic normalization constant
${\cal N}$ must be given by Eq.~(\ref{spanc}).

\end{document}